\documentclass[aps,prb,showpacs,twocolumn]{revtex4-1}
\usepackage{bm}

\usepackage{dcolumn}
\usepackage{color}
\usepackage{amsmath,amssymb,bm, array,tabularx,booktabs,multirow}
\usepackage{longtable}
\usepackage[dvipdfmx]{graphicx}
\usepackage{braket}
\usepackage{txfonts}

\usepackage[normalem]{ulem}

\newcommand{\cum}[1]{\bigl\langle\bigl\langle #1 \bigr\rangle \bigr\rangle}

\begin{document}

\setlength{\extrarowheight}{4pt}

\title{Fluctuation theorem for spin transport at insulating ferromagnetic junctions}

\author{Tetsuya Sato,$^{1}$ Masahiro Tatsuno,$^{1}$ M. Matsuo,$^{2,3,4,5}$ and T. Kato$^{1}$}
\affiliation{
${^1}$Institute for Solid State Physics, The University of Tokyo, Kashiwa 277-8581, Japan\\
${^2}$Kavli Institute for Theoretical Sciences, University of Chinese Academy of Sciences, Beijing 100190, China\\
${^3}$CAS Center for Excellence in Topological Quantum Computation, University of Chinese Academy of Sciences, Beijing 100190, China\\
${^4}$Advanced Science Research Center, Japan Atomic Energy Agency, Tokai 319-1195, Japan\\
${^5}$RIKEN Center for Emergent Matter Science (CEMS), Wako, Saitama 351-0198, Japan\\
}

\date{\today}

\begin{abstract}
General relations for nonequilibrium spin transport at a magnetic junction between a normal metal and a ferromagnetic insulator are derived from the quantum fluctuation theorem.
They include the extended Onsager relations between the spin conductance and the spin-current noise that hold for nonequilibrium states driven by an external current.
These relations, that are valid for a general setup of spin Hall magnetoresistance, provide a comprehensive viewpoint for understanding of unidirectional spin Hall magnetoresistance in insulating ferromagnetic junctions.
\end{abstract}

\maketitle

\section{Introduction}
\label{sec:Introduction}

The fluctuation theorem~\cite{Bochkov1977,Bochkov1979,Evans1993,Esposito2009} has been studied for long time in nonequilibrium statistical mechanics since it can provide a number of powerful results that hold even far from thermodynamic equilibrium.
It connects the transition probability of the forward time evolution with that of the reversed time evolution and leads to general relations for nonlinear transport coefficients.
For example, by applying it to electron transport through a mesoscopic system, various general relations have been proved~\cite{Saito2007,Saito2008,Utsumi2009,Nakamura2010,Nakamura2011} using the framework of full counting statistics~\cite{Levitov1993,Levitov1994,Levitov1996,Levitov2004}.
The fluctuation theorem has also been used to discuss spin-dependent currents in a mesoscopic device coupled with ferromagnetic leads~\cite{Utsumi2010,Tang2018,Wang2015}.

\begin{figure}[tb]
    \centering
    \includegraphics[width=80mm]{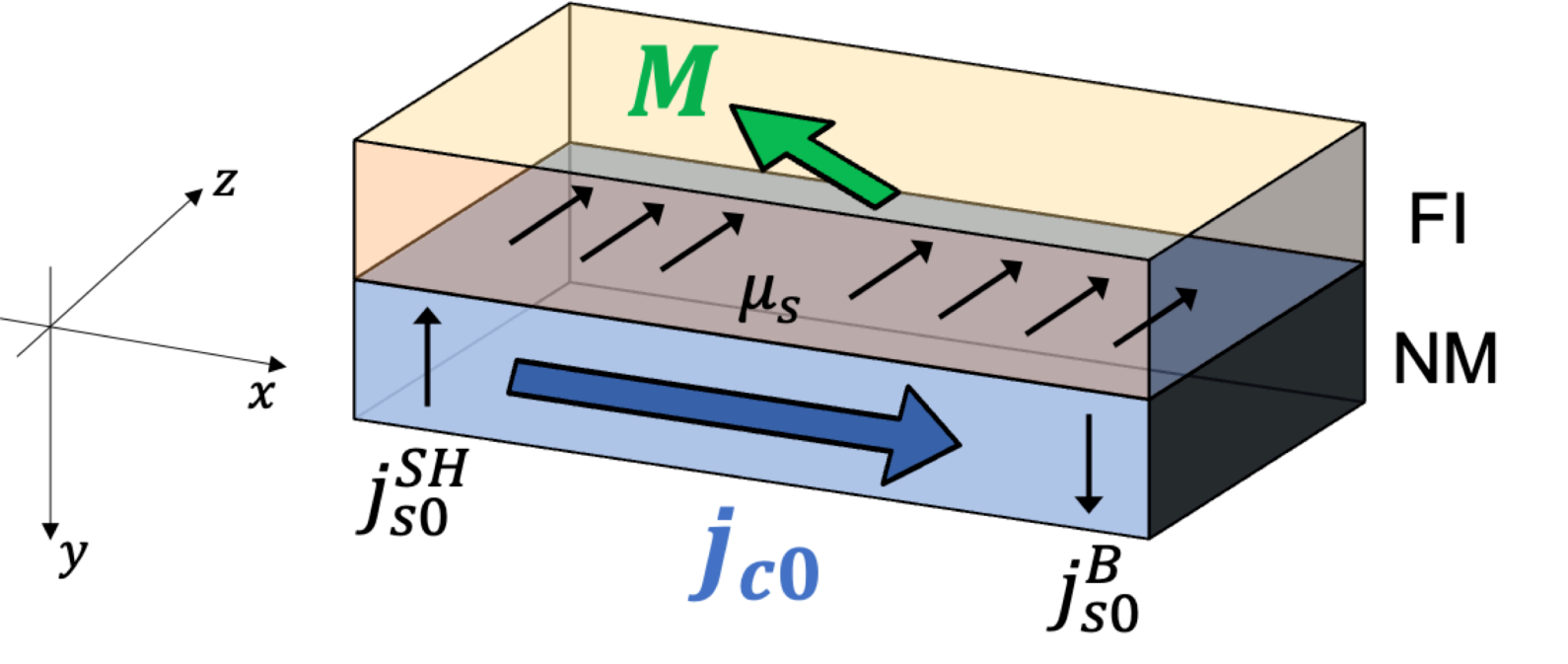}
    \caption{
    Schematic picture of a normal-metal (NM)/ferromagnetic insulator (FI) bilayer system.
    }
    \label{fig:setup}
\end{figure}

In this study, we apply the fluctuation theorem to spin transport in spin Hall magnetoresistance (SMR) that arises in a bilayer system composed of a normal metal (NM) and a ferromagnetic insulator (FI)~\cite{Nakayama2013,Hahn2013,Vlietstra2013,Althammer2013,Meyer2014,Marmion2014,Cho2015,KimSheng2016,Chen2013,Chen2016,Tolle2018}.
The mechanism of SMR is explained as follows~\cite{Nakayama2013,Chen2013,Kato2020} (see Fig.~\ref{fig:setup}).
First, an external current applied to NM induces spin accumulation near the NM/FI interface through the spin Hall effect. 
Next, this spin accumulation causes a backward spin current.
Finally, this backward spin current is converted into a charge current through the inverse spin Hall effect.
This modulated current can be measured by its change when the direction of the magnetization of FI is rotated.
We stress that this mechanism of SMR cannot apply to NM/ferromagnetic metal (FM) junctions in which both charge and spin currents coexist in a complex manner.
In other words, the choice of FI for magnetic junctions makes the situation much simpler, allowing us to consider only spin current at the interface.
It is notable that the use of FI for magnetic junctions has also played a crucial role in experimental study of spin Seebeck effect in concluding that a generated electromotive force under temperature gradient indeed originates from spin current~\cite{Uchida2010}.
This simplification helps us to derive general nonequilibrium relations for spin transport at the NM/FI interface from the fluctuation theorem.

Our work is related to a phenomenon called unidirectional spin Hall magnetoresistance (UMR), which has been observed in NM/FM~\cite{Olejunik2015,Avci2015-1,Avci2015-2,Yin2017,Borisenko2018,Kim2019,Lidig2019,Hasegawa2021} and topological-insulator/FI bilayer systems~\cite{Yasuda2016,Lv2018,Khang2019}.
The UMR is identified by asymmetric contribution of SMR in reverse of the external current and is described by the second-order nonlinear coefficient of the spin current (or the first-order coefficients of the spin conductance) when it is expressed by an expansion with respect to an external current.
Unfortunately, UMR has not been observed so far in NM/FI bilayers~\cite{Avci2015-2}.
However, it is still useful to consider the UMR in NM/FI magnetic junctions as an idealized setup, which allows us to focus only on spin transport without taking care of charge current.
Actually, the UMR for insulating magnetic junctions has been studied in a recent theoretical work~\cite{Sterk2019}.
Our work on the fluctuation theorem for the NM/FI junction system provides a simple explanation for UMR in an idealized situation and will contribute to a comprehensive understanding of SMR.

Let us briefly explain what kind of relations can be derived from the fluctuation theorem. 
Spin accumulation at an interface in the side of the NM is represented by $\mu_s = \mu_\uparrow-\mu_\downarrow$, where $\mu_\uparrow$ and $\mu_\downarrow$ are chemical potentials of conduction electrons for two spin states, respectively.
We note that $\mu_s$ is proportional to the external charge current (see Eqs.~(\ref{mu}) and (\ref{eq:mus0})).
We denote the decay rate of spin angular momenta of the NM at the NM/FI interface as $I_s(\mu_s,{\bm M})$, where ${\bm M}$ is the magnetization of the FI.
The spin conductance, that is defined as $G_s(\mu_s,{\bm M}) \equiv I_s(\mu_s,{\bm M})/\mu_s$, is expanded with respect to $\mu_s$ as
\begin{align}
G_s(\mu_s,{\bm M}) &=  G_s^{(0)}({\bm M})+G_s^{(1)}({\bm M})\frac{\mu_s}{k_BT}+\cdots,
\label{spincurrent}
\end{align}
where $k_B$ is the Boltzmann constant and $T$ is the temperature. 
While SMR is described by the linear coefficient $G_s^{(0)}({\bm M})$, the asymmetric contribution with respect to $\mu_s$, $G_s^{(1)}({\bm M})$, corresponds to the signal of the UMR to be measured in experiments.
Let us also consider spin-current noise at the interface~\cite{Kato2018,Kamra2014}.
The noise power of the spin current, denoted with $S(\mu_s,{\bm M})$, is expanded as
\begin{align}
S(\mu_s,{\bm M})=S^{(0)}({\bm M})+S^{(1)}({\bm M})\frac{\mu_s}{k_BT} + \cdots.
\end{align}
Note that the zeroth-order coefficients satisfy the Onsager relation~\cite{Kato2020},
\begin{align}
S^{(0)}({\bm M}) = 4 \hbar k_{\rm B} T G^{(0)}({\bm M}).
\end{align}
When we rewrite the first-order coefficients as
\begin{align}
\label{pmdef}
G^{(1)}_{s,\pm}({\bm M}) & = G_s^{(1)}({\bm M}) \pm G_s^{(1)}(-{\bm M}), \\
S^{(1)}_{\pm}({\bm M}) & = S^{(1)}({\bm M}) \pm S^{(1)}(-{\bm M}),
\end{align}
we can derive the general relations,
\begin{align}
S_+^{(1)}({\bm M}) & = 12 \hbar k_{\bm B} T G_{s,+}^{(1)}({\bm M}), \label{intromain1} \\ 
S_-^{(1)}({\bm M}) &= 4 \hbar k_{\bm B} T G_{s,-}^{(1)}({\bm M}),
\label{intromain2}
\end{align}
from the quantum fluctuation theorem.
This extended Onsager relation for nonequilibrium states is our main result.
We can also derive general relations for higher-order terms.

This paper is organized as follows.
In Sect.~\ref{sec:UMR}, we theoretically describe SMR by using a combination of spin diffusion theory and spin conductance~\cite{Kato2020}, taking higher-order contributions into account.
We describe the Hamiltonian of this system in Sect.~\ref{sec:ModelHamiltonian} and derive the fluctuation theorem
in Sect.~\ref{sec:FluctuationTheorem}. In Sect.~\ref{sec:ExtendedOnsagerRelation}, we construct the relation between the spin conductance and spin noise. 
We summarize our results in Sect.~\ref{summary}.
The two appendices give a detailed derivation of the fluctuation theorem.

\section{Phenomenological Theory}
\label{sec:UMR}

Before we discuss the fluctuation theorem, we will briefly summarize the phenomenological theory of SMR based on spin diffusion~\cite{Nakayama2013,Chen2013,Kato2020}.
We follow the derivation of Ref.~\onlinecite{Kato2020}.
Consider a NM/FI bilayer system, as shown in Fig.~\ref{fig:setup}.
When an electric field is applied to the NM in the $+x$ direction, a spin current $j_{s0}^{\rm SH}=\theta_{\rm SH}j_{c0}=\theta_{\rm SH}\sigma E_x$ is generated by the spin Hall effect in the $-y$ direction, where $j_{c0}$ is a charge current, $\theta_{\rm SH}$ is the spin Hall angle, $\sigma$ is the electric conductivity, and $E_x$ is the electric field in the $x$ direction. 
This spin current induces spin accumulation near the NM/FI interface that is described by $\mu_s(y)$.
The spatial gradient of $\mu_s(y)$ generates a diffusive backflow spin current $j_s^{\rm B}=-(\sigma/2e)\partial_y \mu_s(y)$, and a steady state is reached by balancing these two spin currents.
For the steady state, we can derive the differential equation,
\begin{align}
\frac{d^2\mu_s(y)}{dy^2} = \frac{\mu_s(y)}{\lambda^2},
\label{diffeq}
\end{align}
where $\lambda$ is the spin diffusion length.
This equation is solved under the boundary condition at the interface of the NM:
\begin{align}
j_s^{\rm B}(y)-j_{s0}^{\rm SH}=
\left\{ \begin{array}{ll}
\displaystyle{-\frac{e}{\hbar/2}\frac{I_s(\mu_s,{\bm M})}{S}}, & (y=0), \\
0, & (y=d_{\rm N}),
\end{array}
\right.
\label{BoundaryCondition}
\end{align}
where $d_{\rm N}$ is the thickness of the NM layer, $S$ is the cross-section area of the NM/FI interface, and $I_s$ is the decay rate of spin angular momenta of the NM at the NM/FI interface.
We define the spin conductance at the interface as 
\begin{align}
G_s(\mu_s,{\bm M}) = \frac{I_s(\mu_s,{\bm M})}{\mu_s} ,
\label{SpinConductance}
\end{align}
where $\mu_s \equiv \mu_s(0)$ is the chemical potential difference at the NM/FI interface. 
The spin conductance is expanded with respect to $\mu_s$ as
\begin{align}
G_s(\mu_s,{\bm M}) = \sum_{m=0}^\infty \frac{1}{m!}G_s^{(m)}({\bm M})\left(\frac{\mu_s}{k_{\rm B}T}\right)^m ,
\end{align}
where $G_s^{(m)}$ is the $m$th-order coefficient of the spin conductance, defined as
\begin{align}
G_s^{(m)}({\bm M}) = (k_{\rm B} T)^m \left. \frac{\partial^m G_s(\mu_s,{\bm M})}{\partial {\mu_s}^m} \right|_{\mu_s=0} .
\end{align}
By solving Eqs.~(\ref{diffeq}), (\ref{BoundaryCondition}), and (\ref{SpinConductance}), we obtain
\begin{align}
\label{mu}
\mu_s &= \frac{\mu_{s0}}{1+g_s(\mu_s,{\bm M})\coth(d_N/\lambda)}, 
\\
\mu_{s0} &=2e\lambda\theta_{SH}E_x\tanh(d_N/2\lambda),
\label{eq:mus0}
\end{align}
where $\mu_{s0}$ is the chemical potential difference in the absence of the NM/FI interface ($G_s=0$) and $g_s(\mu_s,{\bm M})$ is a dimensionless factor defined as
\begin{align}
g_s(\mu_s,{\bm M}) = \frac{4e^2}{\hbar}\frac{G_s(\mu_s,{\bm M})}{\sigma S/\lambda} .
\end{align}
Since $g_s(\mu_s,{\bm M})$ depends on $\mu_s$, Eq.~(\ref{mu}) must be solved self-consistently.
Although $\mu_s$ is not independent of ${\bm M}$ for a fixed $\mu_{s0}$ in general, we can control $\mu_s$ via $\mu_{s0}$, that is proportional to the electric field $E_x$, i.e., to the external current.

Let us first consider the usual explanation of SMR.
If the spin accumulation is sufficiently small, we can approximate the spin conductance as $G_s(\mu_s,{\bm M}) \simeq G_s^{(0)}({\bm M})$.
In this case, we obtain
\begin{align}
& \mu_s = \frac{\mu_{s0}}{1+g_s^{(0)}({\bm M})\coth(d_N/\lambda)} , \\
& g_s^{(0)}({\bm M}) = \frac{4e^2}{\hbar} \frac{G_s^{(0)}({\bm M})}{\sigma S/\lambda}.
\end{align}
The backflow spin current $j_s^{\rm B}$ induces charge current in the $x$ direction by the inverse spin Hall effect and results in a change in resistance that is given as
\begin{align}
\label{rho}
    \frac{\Delta \rho}{\rho}
    & =
    \theta_{SH}^2\frac{g_s(\mu_s,{\bm M})\tanh^2(d_N/2\lambda)}{1+g_s(\mu_s,{\bm M})\coth(d_N/\lambda)}
    \nonumber \\
    & \simeq \theta_{SH}^2\frac{g_s^{(0)}({\bm M})\tanh^2(d_N/2\lambda)}{1+g_s^{(0)}({\bm M})\coth(d_N/\lambda)}.
\end{align}
This magnetoresistance can be measured by changing the direction of the magnetization in the FI through the change in $g_s^{(0)}({\bm M})$.
This is the origin of SMR.

UMR is described as a nonlinear correction to the SMR with respect to $\mu_s$ as follows~\cite{Sterk2019}.
Let us approximate the spin conductance as
\begin{align}
G_s(\mu_s,{\bm M})\simeq G_s^{(0)}({\bm M})+G_s^{(1)}({\bm M})\frac{\mu_s}{k_{\rm B}T},
\end{align}
where the second term on the right-hand side is assumed to be much smaller than the first term.
Accordingly, we obtain
\begin{align}
\label{gs}
g_s(\mu_s,{\bm M}) \simeq g_{s}^{(0)}({\bm M})
+\frac{G_s^{(1)}({\bm M})}{G_s^{(0)}({\bm M})}\frac{g_{s}^{(0)}({\bm M})}{1+g_{s}^{(0)}({\bm M})\coth(d_N/\lambda)}\frac{\mu_{s0}}{k_BT}.
\end{align}
The UMR signal is the difference in $\Delta \rho/\rho$ when the electric current is reversed from the $-x$ direction to the $+x$ direction.
It is calculated as~\cite{Sterk2019}
\begin{align}
\mathcal{U} &= \left.\frac{\Delta \rho}{\rho}\right|_{E_x=E_0}-\left.\frac{\Delta \rho}{\rho}\right|_{E_x=-E_0} \nonumber \\ 
&= \frac{\theta_{SH}^3E_0}{k_BT}\frac{16e^3\lambda^2}{\hbar\sigma S}\frac{\tanh^3(d_N/2\lambda)}{(1+g_{s}^{(0)}({\bm M})\coth(d_N/\lambda))^3}G_s^{(1)}({\bm M}).
\label{expressionU}
\end{align}
From the UMR signal, we can determine $G_s^{(1)}$.
Thus, the linear spin conductance $G_s^{(0)}({\bm M})$ and its correction $G_s^{(1)}({\bm M})$ play an important role in UMR.
The purpose of this work is to relate $G_s^{(1)}({\bm M})$ with the first-order coefficient of the spin-current noise, $S^{(1)}({\bm M})$.

\section{Model}
\label{sec:ModelHamiltonian}

Let us consider the general setup of a NM/FI junction.
We first consider the Hamiltonian for the NM side.
Since spin relaxation in the bulk NM has already been taken into account by spin diffusion theory explained in Sec.~\ref{sec:UMR}, we only need to construct a model for describing conduction electrons near the interface, for which spin relaxation is neglected.
Therefore, we assume that the Hamiltonian the NM side,  $H_{\rm NM}$, conserves the $z$-component of the total spin and is invariant under the time-reversal operation.
Within this restriction, the Hamiltonian $H_{{\rm NM}}$ can be taken arbitrarily.
For example, we can consider the following non-interacting electron system for the NM,
\begin{align}
\label{NMhamiltonian0}
H_{{\rm NM}} = \sum_{{\bm k}\sigma} \xi_{\bm k} c^\dagger_{{\bm k}\sigma} c_{{\bm k}\sigma},
\end{align}
where $c_{{\bm k}\sigma}$ is an annihilation operator of a conduction electron with wavenumber ${\bm k}$ and spin $\sigma$ ($=\uparrow, \downarrow$) and $\xi_{\bm k}$ is a kinetic energy measured from the chemical potential.
One can also include the electron-electron Coulomb interaction.
The spin Hall effect induces spin accumulation on the side of the NM near the interface, which is effectively represented by $\mu_s$ ($=\mu_s(0)$) as
\begin{align}
\label{NMhamiltonian}
{\cal H}_{\rm NM}(\mu_s) &= H_{{\rm NM}} - \mu_s \hat{s}_z, \\
\hat{s}_z &= \frac{1}{2} \sum_{\bm k} (c_{{\bm k}\uparrow}^\dag c_{{\bm k}\uparrow} - c_{{\bm k}\downarrow}^\dag c_{{\bm k}\downarrow}) ,
\end{align}
where $\hat{s}_z$ is the $z$ component of the total spin in the NM.
Here, we note $[{\cal H}_{\rm NM},\hat{s}_z] = 0$.
The Hamiltonian ${\cal H}_{\rm NM}(\mu_s)$ describes the quasi-thermal equilibrium distribution that is reflected in the density matrix and should not be used in the calculation of the time evolution.
Note that no spin current is induced in the absence of spin accumulation and that $\mu_s$ is a driving force for the spin current at the interface.
The Hamiltonian for the FI, denoted as $H_{\rm FI}({\bm M})$, can also be chosen arbitrarily, where ${\bm M}$ is the spontaneous magnetization of the FI.
We only assume that the time reversal operation on $H_{\rm FI}({\bm M})$ gives $H_{\rm FI}(-{\bm M})$~\footnote{One can imagine a realistic situation that the magnetization ${\bm M}$ is controlled by an external magnetic field ${\bm H}$ which is larger than the magnetic anisotropy.
Then, the Zeeman term, that is proportional to $-{\bm H}\cdot {\bm M}$, can be added in the Hamiltonian $H_{\rm FI}$.
In this case, the time reversal operation corresponds to the reverse of the magnetic field ${\bm H}$; $H_{\rm FI}({\bm H}) \rightarrow H_{\rm FI}(-{\bm H})$.}.
This means that our result shown later is valid even in the presence of, e.g., the magnon-magnon and magnon-phonon interactions or the Dzyaloshinskii-Moriya interaction in the FI.

To generate a spin current at the interface between the FI and NM, we need to introduce an interfacial exchange interaction.
Its Hamiltonian, denoted as $H_{\rm int}$, can also be chosen arbitrarily.
For example, in a NM/FI bilayer, we can consider a simple exchange interaction,
\begin{align}
\label{perturbed Hamiltonian}
H_{\rm int} &= \sum_{\langle i,j \rangle} (T_{ij} \hat{S}_i^+ \hat{s}_j^- + T_{ij}^* \hat{S}_i^- \hat{s}_j^+) ,
\end{align}
where $\hat{S}_i^{\pm}$ are spin ladder operators for localized spins in the FI and $\hat{s}_i^{\pm}$ are those for the spins of conduction electrons in the NM, defined as
$\hat{s}_i^+ =  c^\dag_{i,\uparrow}c_{i,\downarrow}$ and $\hat{s}_i^-= c^\dag_{i,\downarrow}c_{i,\uparrow}$, respectively.
This is just a simple example, and one can consider more complex interfacial couplings.

\section{Derivation of Fluctuation Theorem}
\label{sec:FluctuationTheorem}

In this section, we derive the fluctuation theorem following the notation of Refs.~\onlinecite{Esposito2009,Wang2015}.
We assume that the interfacial exchange coupling is switched off for $t \le 0$ so that the initial states of the NM and FI are in thermal equilibrium with the same temperature.
We denote the density matrix at thermal equilibrium at $t=0$ as $\hat{\rho}_0(\mu_s)$.
To express the thermal average, we prepare a complete set of eigenstates for the entire system, as $\ket{\alpha} = \ket{s_z,i} \otimes \ket{j}$, where $\ket{s_z,i}$ is the $i$-th eigen wavefunction of $H_{\rm NM}$ in a sector specified by $s_z$ and $\ket{j}$ is the $j$-th eigen wavefunction of $H_{\rm FI}$.
After the interfacial exchange coupling $H_{\rm int}$ is switched on at $t=0$, we consider the time evolution of the system during $0 \le t \le \tau$ and measure the change of spin $\Delta s$ in the NM by making a projection measurement at $t=\tau$. 
We prepare another complete set of eigenstates for measurement at $t=\tau$, as $\ket{\alpha'} = \ket{s_z',i'} \otimes \ket{j'}$.

The probability function for $\Delta s$ is calculated as
\begin{align}
P(\Delta s_z) &=
\sum_{\alpha,\alpha'} \braket{\alpha|\hat{\rho}_0(\mu_s)|\alpha}
\left|\braket{\alpha'| e^{-iH({\bm M})\tau/\hbar} | \alpha} \right|^2 \delta_{s_z'-s_z,\Delta s_z} ,
\label{DistFunc}
\end{align}
where $H({\bm M})=H_{\rm NM}+H_{\rm FI}({\bm M})+H_{\rm int}$
The initial density matrix $\hat{\rho}_0(\mu_s)$ is described as
\begin{align}
\braket{\alpha|\hat{\rho}_0(\mu_s)|\alpha}
&= Z_0^{-1} e^{-\beta (E_{s_z,i}-\mu_s s_z+E_j)} , 
\end{align}
where $E_{s_z,i}$ and $E_j$ are the eigen energies of $H_{{\rm NM},0}$ and $H_{{\rm FI}}$, respectively, $\beta=1/(k_{\rm B}T)$ is the inverse temperature, and $Z_0={\rm Tr}\, [\hat{\rho}_0(\mu_s)]$ is a partition function.

Next, we consider a time-reversal version of the the probability function.
We take the initial state to be the one that is generated by the time-reversal operation on the thermal equilibrium state.
Such a state is described by $\hat{\rho}_0^{\rm tr}(\mu_s)=\Theta \hat{\rho}_0(\mu_s)\Theta^{-1}$, where $\Theta$ is the time-reversal operator and ${\rm tr}$ indicates a time-reversed state.
We further consider the time evolution by the Hamiltonian $H^{\rm tr}({\bm M})=\Theta H({\bm M}) \Theta^{-1}$ in a period of $\tau$ and perform a projection measurement.
When we express the complete basis set as $\ket{\tilde{\alpha}'}=\Theta \ket{\alpha'}$ for the initial state and $\ket{\tilde{\alpha}}=\Theta \ket{\alpha}$ for the final state, we can define the probability function for the change in the NM spins, $\Delta s_z=(-s_z)-(-s_z')$, under time evolution as
\begin{align}
\label{TRprobability1}
P^{\rm tr}(\Delta s_z) = \sum_{\alpha,\alpha'} \braket{\tilde{\alpha}'|\hat{\rho}_0^{\rm tr}(\mu_s)|\tilde{\alpha}'}
\left|\braket{\tilde{\alpha}|e^{-iH^{\rm tr}({\bm M})\tau/\hbar}|\tilde{\alpha}'}\right|^2 
\delta_{-s_z+s_z',\Delta s_z} .
\end{align}
Here, we assume that the measurement time, $\tau$, is so large that the total energy is conserved (i.e., $E_{\alpha}=E_{\alpha'}$ holds).
Then, it is straightforward to show the relation,
\begin{align}
\label{DetailedFT}
P^{\rm tr}(\Delta s_z)=e^{\beta\mu_s\Delta s_z}P(\Delta s_z).
\end{align}
This relation is called the ``detailed fluctuation theorem'' (for a detailed derivation, see Appendix\ref{app:FTproof}). 
Here, we note that the probability for backward time evolution, $P^{\rm tr}(\Delta s_z)$, can be related to the probability of the forward time evolution after replacing $(\mu_s,{\bm M})$ with $(-\mu_s,-{\bm M})$, because the time-reversal operation corresponds to a reversal of the magnetization and the spin accumulation,
\begin{align}
\hat{\rho}^{\rm tr}_0(\mu_s)&=\hat{\rho}_0(-\mu_s), \\
H^{\rm tr}({\bm M}) &= H(-{\bm M}). 
\end{align}
Therefore, Eq.~(\ref{DetailedFT}) can be rewritten as
\begin{align}
\label{DetailedFT2}
P(\Delta s_z;-\mu_s,-{\bm M})=e^{\beta\mu_s\Delta s_z}P(\Delta s_z;\mu_s,{\bm M})
\end{align}
From $\sum_{\Delta s_z} P(\Delta s_z;-\mu_s,-{\bm M}) = 1$, we also obtain
\begin{align}
\label{FTrelation}
\langle e^{\beta\mu_s\Delta s_z}\rangle_{\mu_s,{\bm M}}= 1,
\end{align}
where $\langle\cdots\rangle_{\mu_s,{\bm M}} = \sum_{\Delta s_z} (\cdots) P(\Delta s_z;\mu_s,{\bm M})$ indicates an average with respect to the probability function $P(\Delta s_z;\mu_s,{\bm M})$.
The $m$th moment of the distribution, $\langle (\Delta s_z)^m \rangle_{\mu_s,{\bm M}}$, generally depends on $\mu_s$
and is expanded with respect to $\mu_s$ as
\begin{align}
\langle (\Delta s_z)^m \rangle_{\mu_s,{\bm M}} = \sum_{l=0}^\infty
\frac{\langle (\Delta s_z)^m \rangle^{(l)}_{{\bm M}}}{l!}
\left(\frac{\mu_s}{k_{\rm B}T}\right)^l,
\label{TaylorExpansion}
\end{align}
where $\langle\cdots\rangle^{(n)}_{\bm M}$ indicates $(k_BT)^n\partial_{\mu_s}^n\langle\cdots\rangle_{\mu_s,{\bm M}}|_{\mu_s = 0}$. By expanding the exponential function in  Eq.~(\ref{FTrelation}) with respect to $\mu_s$, substituting Eq.~(\ref{TaylorExpansion}) into Eq.~(\ref{FTrelation}), and comparing the same orders of $\mu_s$, we obtain
\begin{align}
\label{relation}
\sum_{m=1}^n {}_nC_m\langle(\Delta s_z)^{m}\rangle^{(n-m)}_{\bm M} = 0 .
\end{align}
Concretely, Eq.~(\ref{relation}) can be written for $n=1,2$ and 3 as
\begin{align}
& \langle \Delta s_z\rangle^{(0)}_{\bm M}=0, \\
\label{first}
&\langle \Delta s_z \rangle^{(1)}_{\bm M} + \frac{1}{2} \langle (\Delta s_z)^2\rangle ^{(0)}_{\bm M} = 0, \\
\label{second}
&\langle \Delta s_z\rangle^{(2)}_{\bm M} + \langle (\Delta s_z)^2\rangle^{(1)}_{\bm M}+\frac{1}{3}\langle (\Delta s_z)^3\rangle^{(0)}_{\bm M} = 0. 
\end{align}
In addition, by multiplying $(\Delta s_z)^m$ to Eq.~(\ref{DetailedFT2}) and taking the summation of $\Delta s$, we obtain
\begin{align}
\label{FTtr2}
\langle (\Delta s_z)^{m} \rangle_{-\mu_s,-{\bm M}}=\langle (\Delta s_z)^{m} e^{\beta\mu_s\Delta s}\rangle_{\mu_s,{\bm M}}.
\end{align}
By using the Taylor expansion of Eq.~(\ref{TaylorExpansion}) and
\begin{align}
\langle (\Delta s_z)^m \rangle_{-\mu_s,-{\bm M}} = \sum_{l=0}^\infty
\frac{\langle (\Delta s_z)^m \rangle^{(l)}_{-{\bm M}}}{l!}
\left(-\frac{\mu_s}{k_{\rm B}T}\right)^l,
\label{TaylorExpansion2}
\end{align}
and comparing the same-order terms with respect to $\mu_s$, we can obtain general relations.
For example, for $m=1$, they are obtained from Eq.~(\ref{FTtr2}) as
\begin{align}
\label{third}
-\langle\Delta s_z\rangle^{(1)}_{-{\bm M}}&=\langle\Delta s_z\rangle^{(1)}_{\bm M}+\langle (\Delta s_z)^2\rangle^{(0)}_{\bm M}, \\
\label{fourth}
\langle\Delta s_z\rangle^{(2)}_{-{\bm M}}&=\langle\Delta s_z\rangle^{(2)}_{\bm M}+2\langle(\Delta s_z)^2\rangle^{(1)}_{\bm M} +\langle(\Delta s_z)^3\rangle^{(0)}_{\bm M}.
\end{align}
In a similar way, for $m=2$, we obtain  from Eq.~(\ref{FTtr2}),
\begin{align}
\label{fifth}
\langle (\Delta s_z)^2\rangle^{(0)}_{-{\bm M}}&=\langle(\Delta s_z)^2\rangle^{(0)}_{\bm M},
\\
\label{sixth}
-\langle(\Delta s_z)^2\rangle^{(1)}_{-{\bm M}}&=\langle (\Delta s_z)^2\rangle^{(1)}_{\bm M}+\langle (\Delta s_z)^3\rangle^{(0)}_{\bm M}.
\end{align}
We can obtain relations for higher-order moments and higher-order coefficients with respect to $\mu_s$ by repeating the same procedure.

We rewrite these equations with the cumulants of the distribution functions (see Appendix~\ref{CumulantsRelation}).
Furthermore, we introduce the symmetric and anti-symmetric contributions as
\begin{align}
\cum{\cdots}_\pm^{(n)}=\cum{\cdots}^{(n)}_{\bm M}\pm\cum{\cdots}^{(n)}_{-{\bm M}},
\end{align}
respectively.
Equations~(\ref{first}), (\ref{second}) and (\ref{third})-(\ref{sixth}) are summarized into the following relations:
\begin{align}
\label{main1}
-\cum{\Delta s_z}^{(1)}_+&=\frac{1}{2}\cum{(\Delta s_z)^2}^{(0)}_+
\\
-\cum{\Delta s_z}^{(1)}_-&=\cum{(\Delta s_z)^2}^{(0)}_-=0
\\\label{main2}
-\cum{\Delta s_z}_+^{(2)}&=\frac{1}{3}\cum{(\Delta s_z)^2}_+^{(1)},
\\\label{main3}
-\cum{\Delta s_z}_-^{(2)}&=\cum{(\Delta s_z)^2}_-^{(1)}.
\end{align}

\section{Extended Onsager Relations}
\label{sec:ExtendedOnsagerRelation}

In this section, we describe how the relations obtained from the fluctuation theorem are related to measurable quantities.

\subsection{Spin-current noise}
The average of the spin current, which is defined by the decay rate of NM spins, is written in terms of the first cumulant as 
\begin{align}
I_s(\mu_s,{\bm M})
= -\frac{\hbar}{\tau} \cum{\Delta s_z}_{\mu_s,{\bm M}} .
\label{Isdef}
\end{align}
The noise power of the spin-current noise is defined as
\begin{align}
S(\mu_s,{\bm M})= \frac{1}{\tau} \int_0^{\tau} dt \int_{-\infty}^{\infty} dt' \, &[\langle \Delta \hat{I}_s(t+t') \Delta \hat{I}_s(t)\rangle \nonumber \\
& +\langle \Delta \hat{I}_s(t) \Delta \hat{I}_s(t+t')\rangle] ,
\end{align}
where $\Delta \hat{I}_s(t) = \hat{I}_s(t) - \langle \hat{I}_s(t) \rangle$ and $\hat{I}_s(t)$ is the Heisenberg representation of the spin-current operator.
For a sufficiently large $\tau$, the power noise can be written as
\begin{align}
S(\mu_s,{\bm M}) = \frac{2\hbar^2}{\tau} \cum{(\Delta s_z)^2}_{\mu_s,{\bm M}}.
\label{Sdef}
\end{align}
In section~\ref{sec:Introduction}, we defined the nonlinear coefficients, $G_s^{(m)}({\bm M})$ and $S^{(m)}({\bm M})$ for spin current and its noise as
\begin{align}
G_s(\mu_s, {\bm M}) &= \frac{ I_s(\mu_s,{\bm M})}{\mu_s} = \sum_{m=0}^\infty\frac{G_s^{(m)}({\bm M})}{m!} \left(\frac{\mu_s}{k_{\rm B}T}\right)^m , \\
S(\mu_s,{\bm M}) &= \sum_{m=0}^\infty \frac{S^{(m)}({\bm M})}{m!} \left(\frac{\mu_s}{k_{\rm B}T}\right)^m. 
\end{align}
By comparing these equations with Eqs.~(\ref{Isdef}) and (\ref{Sdef}), we find that
\begin{align}
G_s^{(m)}({\bm M})& =-\frac{\hbar}{(m+1)\tau}\frac{1}{k_BT}\cum{\Delta s_z}^{(m+1)}_{\bm M}, \\
S^{(m)}({\bm M})&= \frac{2\hbar^2}{\tau} \cum{(\Delta s_z)^2}^{(m)}_{\bm M}.
\end{align}
Here, we define the symmetric and anti-symmetric parts of the spin conductance and spin-current noise as
\begin{align}
G^{(n)}_{s,\pm}({\bm M}) & =G^{(n)}_s({\bm M})\pm G^{(n)}_s(-{\bm M}), \\
S^{(n)}_\pm({\bm M}) & =S^{(n)}({\bm M})\pm S^{(n)}(-{\bm M}).
\end{align}
Note that these quantities are easily measured in experiments by rotating the magnetization of the FI with an external magnetic field.
Using Eq.~(\ref{main1}), we obtain
\begin{align}
\label{0th}
& S^{(0)}_+({\bm M})=4\hbar k_BT G_{s,+}^{(0)}({\bm M}), \\ 
& S^{(0)}_-({\bm M})=G_{s,-}^{(0)}({\bm M})=0.
\end{align}
The first relation is the Onsager relation that holds for a system in thermal equilibrium.

Nontrivial relations can be obtained from  Eqs.~(\ref{main2}) and (\ref{main3}).
By rewriting these equations with the spin conductance and the spin-current noise, we obtain
\begin{align}
S^{(1)}_+({\bm M}) &= 12\hbar k_BTG_{s,+}^{(1)}({\bm M}), \label{main1a} \\
S^{(1)}_-({\bm M}) &= 4\hbar k_BTG_{s,-}^{(1)}({\bm M}).  \label{main2a}
\end{align}
These relations are our main result.

As shown in Sect.~\ref{sec:UMR}, $G_{s}^{(1)}({\bm M})$ is related to the UMR in the NM/FI bilayer system. 
These relations indicate that the signal of UMR in NM/FI junctions is directly related to the first-order coefficient of the spin-current noise with respect to $\mu_s$.
These nontrivial relations, which can be called the extended Onsager relations, are expected to be verified by careful observation of UMR and spin-current noise in the presence of external current in the NM.
We note that our result is derived for a simplified setup, i.e., insulating magnetic junctions, which allow us to focus on spin transport at the interface.
In the present simple setup, our result indicates that the UMR, as well as more higher contribution with respect to the external current, may be understood in a unified framework of nonequilibrium statistics.
This helpful viewpoint will contribute to further understanding on the UMR.

\subsection{Higher-order cumulant}

We can also derive general relations including higher-order coefficients and higher-order cumulants.
We define the third cumulant of the spin current as
\begin{align}
C(\mu_s,{\bm M}) = -\frac{\hbar^3}{\tau}\cum{\Delta s_z^3}_{\mu_s,{\bm M}},
\end{align}
and its Taylor expansion as
\begin{align}
C(\mu_s,{\bm M}) &= \sum_{m=0}^\infty \frac{C^{(m)}({\bm M})}{m!} \left(\frac{\mu_s}{k_{\rm B}T}\right)^m, \\
C^{(m)}({\bm M}) & = - \frac{\hbar^3}{\tau} \cum{(\Delta s_z)^3}_{{\bm M}}^{(m)}. \end{align}
We further define the symmetric and anti-symmetric parts as 
\begin{align}
C_\pm^{(m)}({\bm M}) &= C^{(m)}({\bm M}) \pm C^{(m)}(-{\bm M}),
\end{align}
respectively.
Then, from the fluctuation theorem, we can derive the following relations:
\begin{align}
C^{(0)}_-({\bm M}) & = 0, \\
C^{(0)}_+({\bm M}) & =\hbar S^{(1)}_+({\bm M})=12\hbar^2 k_BT G_{s,+}^{(1)}({\bm M}), \\
C^{(1)}_-({\bm M}) &=\frac{\hbar}{2}S^{(2)}_-({\bm M})=6\hbar^2 k_{\rm B}TG_{s,-}^{(3)}({\bm M}), \\
C^{(1)}_+({\bm M}) & = \frac{3\hbar}{2} S^{(2)}_+({\bm M})
- 6 \hbar^2 k_{\rm B}T G_{s,+}^{(3)}({\bm M}).
\end{align}
By repeating a similar procedure, we can derive more relations for higher-order coefficients and higher-order cumulants.

\section{Summary}
\label{summary}

We applied the quantum fluctuation theorem to a typical setup for the SMR, i.e., to a bilayer system composed of a NM and a FI.
We derived general relations on the average of the spin current, spin-current noise, and higher-order cumulants of the spin current.
In particular, we showed that the first-order correction of the spin conductance with respect to $\mu_s$ is related to that of the spin-current noise (see Eqs.~(\ref{main1a}) and (\ref{main2a})).
These relations, which should be called the Onsager relations extended to nonequilibrium states, can be verified in experiments through a measurement of the UMR in insulating magnetic junctions.
We expect that our result will contribute to clear understanding of the SMR and the UMR.

\section*{ACKNOWLEDGMENTS}

We thank Y. Ominato for discussions. T. K. acknowledges support from the Japan Society for the Promotion of Science (JSPS KAKENHI Grants No. 20K03831).
M. M. is supported by JSPS KAKENHI for Grants (Nos. 20H01863 and 21H04565) and the Priority Program of the Chinese Academy of Sciences, Grant No. XDB28000000.

\appendix

\section{Derivation of the Fluctuation Theorem}
\label{app:FTproof}

In this appendix, we derive the detailed fluctuation theorem, Eq.~(\ref{DetailedFT}).
We start with the distribution function, $P(\Delta s_z)$, defined in Eq.~(\ref{DistFunc}).
By inserting $\Theta^{-1} \Theta = 1$, we obtain
\begin{align}
P(\Delta s_z) &= \sum_{\alpha,\alpha'} \delta_{s_z'-s_z,\Delta s_z}
\braket{\tilde{\alpha}|\hat{\rho}^{\rm tr}_0(\mu_s)|\tilde{\alpha}}
|\braket{\tilde{\alpha}'|e^{iH^{\rm tr}({\bm M})t}|\tilde{\alpha}}|^2,
\label{FTproof1}
\end{align}
where $\ket{\tilde{\alpha}} = \Theta \ket{\alpha}$ and $\ket{\tilde{\alpha}'} = \Theta \ket{\alpha'}$.
The density matrix $\hat{\rho}^{\rm tr}_0(\mu_s)$ is calculated as
\begin{align}
\hat{\rho}^{\rm tr}_0(\mu_s) &= 
\Theta \hat{\rho}_0 (\mu_s)\Theta^{-1} \nonumber \\
&= \frac{1}{Z_0} \exp( -\beta {\cal H}^{\rm tr}(\mu_s,{\bm M})), \\
{\cal H}^{\rm tr}(\mu_s,{\bm M}) &= \Theta (H_{\rm NM} -\mu_s \hat{s}_z + H_{\rm FI}({\bm M})) \Theta^{-1} \nonumber \\
&= H_{\rm NM}^{\rm tr} + \mu_s \hat{s}_z + H_{\rm FI}^{\rm tr}({\bm M}).
\end{align}
Because $[H_{\rm NM},\hat{s}_z] = 0$, the element of the density matrix in the sum of Eq.~(\ref{FTproof1}) can be rewritten as
\begin{align}
\braket{\tilde{\alpha}|\hat{\rho}^{\rm tr}_0(\mu_s)|\tilde{\alpha}}
&= \frac{1}{Z_0} \exp\left[ -\beta( E_{s_z,i}+E_j-\mu_s s_z ) \right] 
\nonumber \\
&= e^{-\beta \mu_s\Delta s_z} \braket{\tilde{\alpha}'|\rho_0^{\rm tr} (\mu_s)|\tilde{\alpha}'}.
\end{align}
In the second equation we have used $s_z = s_z'-\Delta s$, that holds in the presence of the delta function $\delta_{s_z'-s_z,\Delta s_z}$, and energy conservation $E_{s_z,i}+E_{j}=E_{s_z',i'}+E_{j'}$.
Finally, we obtain the detailed fluctuation theorem,
\begin{align}
P(\Delta s_z) &= e^{-\beta \mu_s\Delta s_z} \sum_{\alpha,\alpha'}
\delta_{s_z'-s_z,\Delta s_z}
\braket{\tilde{\alpha}'|\hat{\rho}_0^{\rm tr}(\mu_s) |\tilde{\alpha}'}
\nonumber \\
& \hspace{10mm} \times |\braket{\tilde{\alpha}|e^{-iH^{\rm tr}({\bm M})t/\hbar}|\tilde{\alpha}'}|^2 \nonumber \\
&= e^{-\beta \mu_s\Delta s_z} P^{\rm tr}(\Delta s_z).
\end{align}

\section{Relation between cumulants and moments}
\label{CumulantsRelation}

The cumulants of the distribution function are defined from the moments as
\begin{align}
\cum{\Delta s_z}_{\bm M} &= \langle \Delta s_z \rangle_{\bm M}, \\ 
\cum{(\Delta s_z)^2}_{\bm M} &= \langle (\Delta s_z)^2 \rangle_{\bm M}- \left[\langle  \Delta s_z \rangle_{\bm M}\right]^2, \\ 
\cum{(\Delta s_z)^3}_{\bm M} &= 
\langle (\Delta s_z)^3 \rangle_{\bm M}
- 3 \langle (\Delta s_z)^2 \rangle_{\bm M} \langle \Delta s_z \rangle_{\bm M} \nonumber \\
& \hspace{5mm} + 2 \left[\langle \Delta s_z \rangle_{\bm M}\right]^3 , 
\end{align}
and so on.
We also define the coefficients in their Taylor expansion as
\begin{align}
\cum{(\Delta s_z)^m}_{\bm M} =  \sum_{l=0}^\infty
\frac{\cum{(\Delta s_z)^m}^{(l)}_{{\bm M}}}{l!}
\left(\frac{\mu_s}{k_{\rm B}T}\right)^l,
\label{TaylorExpansionCum}
\end{align}
where $\langle \langle\cdots\rangle\rangle^{(n)}=(k_BT)^n\partial_{\mu_s}^n\langle \langle\cdots\rangle \rangle|_{\mu_s=0}$.
Using the relations
\begin{align}
\cum{\Delta s_z}^{(0)}_{\bm M} &= \langle\Delta s_z\rangle^{(0)}_{\bm M} = 0, \\
\cum{(\Delta s_z)^2}^{(0)}_{\bm M} &=\langle (\Delta s_z)^2\rangle^{(0)}_{\bm M}-\left[\langle\Delta s_z\rangle^{(0)}_{\bm M} \right]^{2} =\langle (\Delta s_z)^2\rangle^{(0)}_{\bm M},
\\
\cum{(\Delta s_z)^2}^{(1)}_{\bm M} &=\langle (\Delta s_z)^2\rangle^{(1)}_{\bm M}- 2\langle\Delta s_z\rangle^{(0)}_{\bm M} \langle\Delta s_z\rangle^{(1)}_{\bm M} \nonumber \\
&=\langle(\Delta s_z)^2\rangle^{(1)}_{\bm M} ,
\\
\cum{(\Delta s_z)^3}^{(0)}_{\bm M}&=\langle\Delta s_z^3\rangle^{(0)}_{\bm M}-3\langle\Delta s_z^2\rangle^{(0)}_{\bm M} \langle\Delta s_z\rangle^{(0)}_{\bm M}+2\left[\langle\Delta s_z\rangle^{(0)}_{\bm M} \right]^{3} \nonumber \\
&=\langle (\Delta s_z)^3\rangle^{(0)}_{\bm M},
\end{align}
we can rewrite Eqs.~(\ref{first}) and (\ref{second}) as
\begin{align}
& \langle\langle\Delta  s_z\rangle\rangle^{(1)}_{\bm M}+\frac{1}{2}\langle\langle \Delta s_z^2\rangle\rangle^{(0)}_{\bm M} =0,
\label{FT1}
\\\label{FT2}
& \langle\langle\Delta s_z\rangle\rangle^{(2)}_{\bm M}+\langle\langle\Delta s_z^2\rangle\rangle^{(1)}_{\bm M}+\frac{1}{3}\langle\langle\Delta s_z^3\rangle\rangle^{(0)}_{\bm M} =0.
\end{align}
In a similar way, we can rewrite
Eqs.~(\ref{third})-(\ref{sixth}) as
\begin{align}
-\cum{\Delta s_z}^{(1)}_{-{\bm M}} & =\cum{\Delta s_z}^{(1)}_{{\bm M}}
\label{tr2} +\cum{\Delta s_z^2}^{(0)}_{\bm M},\\
\cum{\Delta s_z}^{(2)}_{-{\bm M}} & =\cum{\Delta s_z}^{(2)}_{\bm M}+2\cum{\Delta s_z^2}^{(1)}_{\bm M} \nonumber \\
&\hspace{5mm} +\cum{\Delta s_z^3}^{(0)}_{\bm M},
\\
\cum{\Delta s_z^2}^{(0)}_{-{\bm M}} & =\cum{\Delta s_z^2}^{(0)}_{\bm M},
\\\label{2tr1}
-\cum{\Delta s_z^2}^{(1)}_{-{\bm M}} &=\cum{\Delta s_z^2}^{(1)}_{\bm M}+\cum{\Delta s_z^3}^{(0)}_{\bm M} ,
\end{align}
respectively.

\bibliography{FTSpin.bib}
\end{document}